\documentclass[%
 reprint,
 groupedaddress,
 showpacs, 
 noshowkeys,
 amsmath,amssymb,
 aps,
 prl,
]{revtex4-1}

\usepackage{graphicx}
\usepackage{dcolumn}
\usepackage{bm}
\usepackage{hyperref}

\usepackage[cdot,mediumqspace,amssymb]{SIunits} 
\usepackage[autolanguage]{numprint} 

\usepackage{ifthen}
\usepackage{ifpdf}
\ifpdf
    \graphicspath{{Figures/PDF/}}
\else
    \graphicspath{{Figures/EPS/}}
\fi

\usepackage{cleveref}
\usepackage{color}
\newcommand{\DB}[1]{{\color{black} #1 }}
\newcommand{\Lisa}[1]{{\color{black} #1 }}
\newcommand{\map}[1]{{\color{black} #1}}
\newcommand{\KD}[1]{{\color{black} #1 }}

\begin{document}

\title{Extraction of Force-Chain Network Architecture in Granular Materials Using Community Detection}

\author{Danielle S. Bassett$^{1,2,*}$, Eli T. Owens$^{3,4}$, Mason A. Porter$^{5,6}$, M. Lisa Manning$^{7}$, Karen E.
Daniels$^{3}$}
\affiliation{$^{1}$ Department of Bioengineering, University of Pennsylvania, Philadelphia, PA 19104, USA;\\
$^2$Department of Electrical Engineering, University of Pennsylvania, Philadelphia, PA 19104, USA;\\
$^3$Department of Physics, North Carolina State University, Raleigh, NC 27695, USA;\\
$^4$Department of Physics, Presbyterian College, Clinton, SC 29325, USA;\\
$^5$Oxford Centre for Industrial and Applied Mathematics, Mathematical Institute, University of Oxford, Oxford OX2 6GG, UK;\\
$^6$CABDyN Complexity Centre, University of Oxford, Oxford, OX1 1HP, UK;\\
$^7$Department of Physics, Syracuse University, Syracuse, NY 13244, USA\\
$^*$Corresponding author. E-mail address: dsb@seas.upenn.edu}

\date{\today}



\begin{abstract}

Force chains form heterogeneous physical structures that can constrain the mechanical stability and acoustic transmission of granular media. However, despite their relevance for predicting bulk properties of materials, there is no agreement on a quantitative description of force chains. Consequently, it is difficult to compare the force-chain structures in different materials or experimental conditions. To address this challenge, we treat granular materials as spatially-embedded networks in which the nodes (particles) are connected by weighted edges that represent contact forces. We use techniques from community detection, which is a type of clustering, to find \map{sets} of closely connected particles.  By using a \emph{geographical null model} that is constrained by the particles' contact network, we extract chain-like structures that are reminiscent of force chains. We propose three diagnostics to measure these chain-like structures, and we demonstrate the utility of these diagnostics for identifying and characterizing classes of force-chain network architectures in various materials. To illustrate our methods, we describe how force-chain architecture depends on pressure for \DB{two very different types of packings: \map{(1) ones} derived from laboratory experiments and \map{(2) ones} derived from \KD{idealized, }numerically-generated frictionless packings.} By resolving individual force chains, we quantify statistical properties of force-chain shape and strength, which are potentially crucial diagnostics of bulk properties (including material stability). These methods facilitate quantitative comparisons between \DB{different} particulate systems, regardless of whether they are measured experimentally or numerically.

\end{abstract}


\maketitle



\section{Introduction \label{sec:introduction}}

\begin{figure}
\begin{center}
\includegraphics[width=0.8\linewidth]{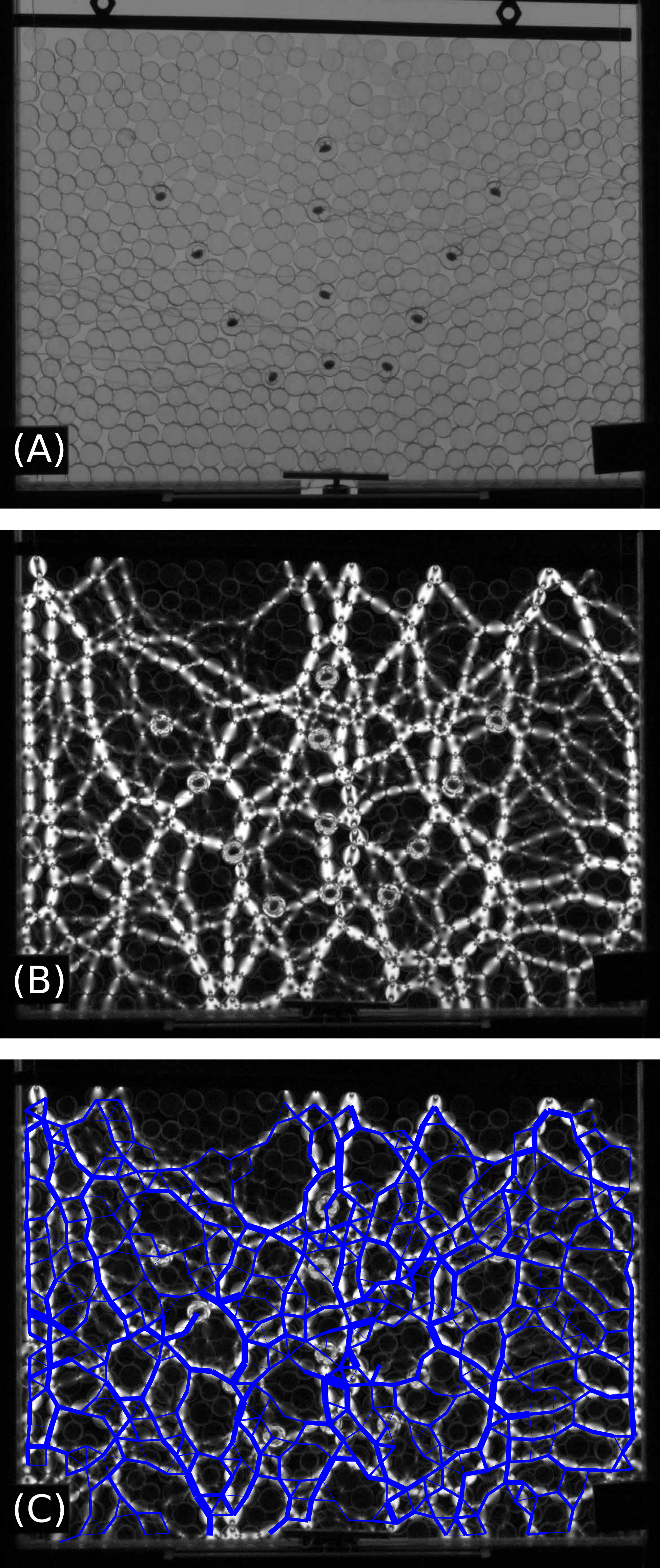}
\end{center}
\caption[]{
(Color Online) We represent granular materials as spatially-embedded networks \cite{barth2011} whose nodes (particles) are connected by weighted
edges that represent contact forces. \emph{(A)} Image of a 2D vertical
aggregate of photoelastic disks that are confined in a single layer by a pressure of $P \approx 6.7 \times 10^{-4}$E. Several particles are embedded with a piezoelectric sensor, for which the wires are visible.
\emph{(B)} The internal stress pattern in the photoelastic particles
manifests as a network of force chains. \emph{(C)} The blue line segments show the edges of a
weighted graph, which we determine from image processing and overlay on the image from panel \emph{(B)}. An edge between two particles (i.e., nodes) exists if the two
particles are in physical contact with each other; the forces between
particles give the weights of the edges. \DB{Note: the orientation of the packings is larger horizontal coordinates to the right, and larger vertical coordinates upward.}
\label{f:granexp}}
\end{figure}

Particulate matter comes in many forms: it ranges from frictionless emulsions and frictional granular materials to bonded composites and biological cells. A long-known hallmark of particulate systems is the heterogeneous distribution of inter-particle forces within them \cite{Dantu-1957-CEM,Liu-1995-FFB, Radjai1996,Mueth1998,Howell1999,Majmudar2005}---a phenomenon that has come to be known as \emph{force chains}. The term ``force chain'' arises from the appearance---particularly in two-dimensional (2D) systems---of chains of particles that transfer force from one \map{particle} to another along the chain. In Fig.~\ref{f:granexp}, for example, the eye readily focuses on the backbone that arises from the set of bright particles. These particles experience forces that are larger than average, and one can see such forces by using photoelastic particles and placing them between crossed polarizers. In both two and three dimensions, it is increasingly common to measure inter-particle forces in a variety of systems -- including 
frictional grains \cite{Majmudar2005,Saadatfar2012}, frictionless grains \cite{Mukhopadhyay2011}, and frictionless emulsions \cite{Brujic2003, Zhou2006a, Desmond2013}. This opens the door for achieving a better understanding of force chains in a variety of systems.

It is known that force chains are important for resisting shear \cite{Howell1999} and directing sound propagation \cite{Makse-1999-WEM,Hidalgo2002,Somfai2005,Owens2011}, but little is known about which properties of these chains are universal to particulate systems and which are sensitive to details (such as friction, adhesion, boundary conditions, and body forces like gravity). Improved methods for automatically identifying force chains and quantifying their specific properties (e.g. size and shape) would yield a deeper understanding of how they impact mechanical properties of particulate systems.

Although the presence of force chains is a generic feature of particulate systems, the term ``force chain'' is often used colloquially, and the field still lacks a quantitative definition of the term. Recently, several techniques have been proposed that aim to identify which subset of \map{particles} form a ``force-chain network.''
Peters et al. \cite{Peters2005} calculated force chains in low friction by \map{demanding} two requirements: particles must occur in a ``quasi-linear'' arrangement, and they must contain a concentrated stress. They used an algorithm that requires a choice of a threshold value for each of the two conditions, and their notion does not allow force chains to branch. Using discrete-element simulations with a flat-punch geometry, Peters et al. \cite{Peters2005} obtained force chains that \map{contain} approximately half of the particles in a system.  Additionally, \map{they reported that the lengths of these chains satisfy} an approximately exponential distribution. Arevalo et al. \cite{Arevalo2010} examined polygonal structures that underly a force-chain network, which they determined by selecting all inter-particle forces above some threshold. In their numerical simulations, they \map{observed} that triangular structures dominate the network near a critical packing fraction. More recently, topological tools such 
as computational homology have been used to address the question of how to determine the structure of force chains \cite{Kondic2012,kramar2014,maza2014}.
For example, \map{using numerical simulations,} Kondic et al. \cite{Kondic2012} were able to distinguish between frictional and frictionless packings \map{via} a topological invariant known as the $0^{\mathrm{th}}$ Betti number \cite{edels2010}. They demonstrated that the $0^{\mathrm{th}}$ Betti number \map{changes} with packing density, and \map{they used} a force threshold to isolate strong forces from weak forces.

\map{The above investigations have} been informative, but they also share a common viewpoint that the strongest inter-particle forces form the backbone of a particulate system. However, even a linear chain of strong particles would not be stable against buckling without the participation of the particles that lie alongside them \cite{Radjai1998}. Therefore, it is necessary to develop techniques that do not include a minimum threshold force to be able to consider a \map{particle to be part of a network}.

The purpose of the present paper is to use an approach based on network science \cite{Newman2010} to develop an example of \map{an} appropriate method. Network science provides a powerful set of tools to represent complex systems by focusing not only on the components of such systems, but also on the interactions among those components. A network representation is particularly appropriate for particulate systems, where particles (network nodes) are connected to one another by contact forces (network edges).

In Fig.~\ref{f:granexp}, we show an example network, which we constructed from laboratory experiments.  Network approaches in granular materials have already had several success in describing the dynamics of granular materials. For example, prior studies have investigated spatial patterns in the breaking of edges in a (binary) contact network under shear \cite{Herrera2011b, Tordesillas2010, Walker-2010-TED,Slotterback2012} and the influence of network topology on acoustic propagation \cite{Bassett2012}. The latter study established the importance of using weighted contact networks to take into account the strength of inter-particle forces. Several other papers have also recently contributed to this line of inquiry using a variety of different \map{network-based} approaches \cite{Lopez2013,arevalo2013, ant2014,Navakas2014}. 

In the present paper, we use network representations to build a set of practical methods to (1) automatically extract force chains from force networks (without the use of thresholding) and (2) define a set of three scalar quantities that one can use to characterize and classify different particulate materials. In Section \ref{sec:Methods}, we present two commonly-used granular systems \map{(which are rather different from each other)} in which it is possible to measure inter-particle forces. These two case studies---one in a ``wet'' laboratory and the other based on a standard computational model for jammed packings---provide a basis for describing the utility of our new technique. In Section \ref{sec:forcechain}, we describe our method to automatically extract the force-chain structure from a force network.

\DB{Our method uses community detection, which is a form of clustering for networks \cite{Porter2009,Fortunato2010}. In contrast to previous uses of community detection in particulate systems \cite{zohar2011,Bassett2012}, our method incorporates a \emph{geographical null model} to account for spatial embeddedness.} \DB{This geographical null model \DB{makes it possible to} to identify clusters of particles that are more densely interconnected \DB{via} strong contact forces than expected given their geographical proximity.} We then calculate a \emph{gap factor} that allows \map{one} to optimize the resolution at which the detected communities are maximally branched. \DB{One can use the gap factor to quantify similarity of a community's geometry to that expected in a force chain. The gap factor is larger when communities are more branch-like and thus more similar to the expected geometry of a force chain; it is smaller when communities are either more compact or more string-like, and thus less similar to the expected geometry of a force chain. In addition to} the gap factor, we use two other diagnostics---size and network force---to help characterize particulate structures. (We define \map{the three} diagnostics in Section \ref{props}.) We then apply our methodology to force networks that we \map{measure} in two \DB{very different} case studies.  We demonstrate the utility of our diagnostics for (1) quantifying changes in force-chain structure as a function of the confining pressure applied to a granular material and (2) facilitating the comparison and classification of force-chain network architectures across different media. We conclude and discuss several implications of our work in Section \ref{sec:Discussion}.


\section{Methods \label{sec:Methods}} 

To better understand which features of force chains are universal and to demonstrate that our methodology for identifying force chains is insightful for particulate systems in general, we compare and contrast results from two \DB{very different} case studies: granular experiments and frictionless simulations. In both situations, the particles are confined in two dimensions, and they each consist of packings of bidisperse disks that interact with each other in a Hertzian-like manner. Both cases are jammed under constant pressure and use approximately the same number of particles (600). The key differences between the two systems are (1) the presence of friction and gravitational pressure in the experiments and (2) the use of periodic boundary conditions in the simulations. \DB{Additional differences include a different fabric tensor due to different initial conditions and the fact that laboratory experiments have a slight fine-scale polydispersity for particles of the ``same'' size in addition to 
the coarse-scale bidispersity.} For each of the experiments and simulations, we measure force-contact networks for (approximately) the same \map{seven} different values of confining pressure.


\subsection{Granular Experiments}

We perform experiments on a vertical 2D granular system of bidisperse disks that are confined between two sheets of Plexiglas. The particles are $6.35$~mm thick, their diameters are $d_1 = 9$~mm and $d_2=11$~mm (which yields a diameter ratio of approximately 1.22), and they are cut from Vishay PSM-4 photoelastic material to provide measurements of the internal forces. These particles have an elastic modulus of $E=4$~MPa. We produce new configurations by rearranging the particles by hand. We increase the pressure by placing additional brass weights on the top surface of the packing. The values of pressure, which we report in units of the elastic modulus $E$ (recall that the configuration is two-dimensional), are
$2.7 \times 10^{-4}E$, $4.1 \times 10^{-4}E$, $6.7 \times 10^{-4}E$,
$1.1 \times 10^{-3}E$, $2.2 \times 10^{-3}E$, $3.8 \times 10^{-3}E$, and $5.9 \times 10^{-3}E$. \DB{Particles are constrained by vertical walls to prevent deformation due to these pressures from occurring in the direction perpendicular to the loading direction. Such constraints on deformation can influence the shape of force chains, which tend to form in the direction of the major principal stress \cite{Tordesillas2010,Majmudar2005,Radjai2009}.} See Refs.~\cite{Owens2011,Bassett2012,Owens2013} for additional details about the experiments.

For each of 21 particle configurations and the \map{seven} values of pressure, we compute particle positions and forces using two high-resolution pictures of the system.  We use one image, which we take without the polarizers, to determine the particle positions and contacts. (See \cite{Puckett2013} for a description of the technique.) \map{We take particles to be in contact if the force between them is measurable by our photoelastic calculations.} Using a second image that we take with the polarizers, we then determine the particle contact forces by solving the inverse photoelastic problem \cite{Puckett2013}.


\subsection{Frictionless Simulations}

We perform numerical simulations of a 50:50 ratio of bidisperse frictionless disks.  We use a diameter ratio of $1.22$, and particles interact via a Hertzian potential in a box with periodic boundary conditions in both directions and zero gravity~\cite{OHernCG, OHern2003,Manning2011}. This model has been well-studied and it is significantly different from our experimental system \map{with respect to} friction, gravity, and \map{the} fixed boundaries.  We generate mechanically-stable packings via a standard conjugate-gradient method~\cite{OHernCG}. We then perform simulations for a fixed packing fraction and volume, and we analyze 20 mechanically-stable packings at each packing fraction $\phi$.  We choose the seven values of the packing fraction so that the mean pressure $\overline{p}$ at that packing fraction~\footnote {For finite systems, there isn't a bijection between pressure and packing fraction. Because the packing fraction is fixed in these simulations, there is some variance (of roughly $9 \times 10^{
-5}$) in the pressure 
between different packings.} matches the ones in the experiments: $[\phi, \overline{p}] = [0.8499, 3 \times 10^{-4} E]$,
$[0.8521, 4 \times 10^{-4} E]$,
$[0.8560, 7 \times 10^{-4} E]$,
$[0.8621, 11 \times 10^{-4} E]$,
$[0.8760, 22 \times 10^{-4} E]$,
$[0.8927, 38 \times 10^{-4 }E]$, and
$[0.9106, 59 \times 10^{-4} E]$,
where the modulus $E$ is defined as the energy scale for the Hertzian interaction $\epsilon$ divided by the mean Vorono\"i area of a particle in the packing. The \map{smallest} value of $\phi$ provides a data point for a jammed packing that is less dense than what is accessible in our experiments.


\section{Force-Chain Extraction and Characterization \label{sec:forcechain}}

\subsection{Force-Chain Extraction}

To locate the force chains, we begin by recording which particles are in contact with (and exert force on) which other particles. In network language, we represent each particle as a node, and we represent each inter-particle contact as an edge whose weight is given by the magnitude of the force at that contact.
We thereby construct a force-weighted contact network ${\bf W}$ from a list of all inter-particle forces.
If particle $i$ and $j$ are in contact, \DB{then $W_{ij} = f_{ij}/\mathrm{mean}(f)$,
where $f_{ij}$ is the normal force between them.}
If two particles are not in contact, then $W_{ij} = 0$. In addition, we let $W_{ii} = 0$.  We also construct an unweighted (i.e., binary) matrix ${\bf B}$ whose elements are
\begin{equation*}
	B_{ij} = \begin{cases}
		1\,, & W_{ij} \neq 0\,, \\
				0\,, & W_{ij} = 0\,.
	\end{cases}
\end{equation*}	
The matrix ${\bf B}$ is often called an ``adjacency matrix'' \cite{Newman2010}, and the matrix ${\bf W}$ is \map{often called a} ``weight matrix.'' 

To obtain force chains from ${\bf W}$, we want to determine sets of particles for which strong inter-particle forces occur amidst densely connected sets of particles. We can obtain a solution to this problem via ``community detection'' \cite{Porter2009,Fortunato2010,Newman2012}, in which we seek \map{sets} of densely connected nodes called ``modules'' or ``communities.''  A popular way to identify communities in a network is by maximizing \map{a quality function known as \emph{modularity}} \DB{with respect to the assignment of particles to \DB{sets called ``communities.'' Modularity $Q$ is defined as}}
\begin{equation}\label{eq:one}
	Q = \sum_{i,j} [W_{ij} - \gamma P_{ij}] \delta(c_{i},c_{j})\,,
\end{equation}
where node $i$ is assigned to community $c_{i}$, node $j$ is assigned to
community $c_{j}$, the Kronecker delta $\delta(c_{i},c_{j})=1$ if and only if $c_{i} =
c_{j}$, the quantity $\gamma$ is a resolution parameter, and
$P_{ij}$ is the expected weight of the edge that connects node $i$ and node $j$
under a specified null model. 

\DB{One can use the maximum value of modularity to quantify the quality of a partition of a force network into sets of particles that are more densely interconnected by strong forces than expected under a given null model. The resolution parameter $\gamma$ provides a means of probing the organization of inter-particle forces across a range of spatial resolutions. To provide some intuition, we note that a perfectly hexagonal packing with non-uniform forces should still possess a single community for small values of $\gamma$ and should consist of a collection of single-particle (i.e., singleton) communities for large values of $\gamma$. At intermediate values of $\gamma$, we expect maximizing modularity to yield a roughly homogeneous assignment of particles into communities of some size (i.e., number of particles) between $1$ and the total number of particles. (The exact size depends on the value of $\gamma$.) The strongly inhomogeneous community assignments that we observe in the laboratory and numerical packings (see Section \ref{sec:measure}) are a direct consequence of the disorder in the packings.}

\DB{An important choice in maximizing modularity optimization is the null model $P_{ij}$ \cite{Newman2006PRE,Bassett2013Robust}.} The most common null model for modularity optimization is the Newman-Girvan (NG) null model \cite{NG2004,Newman2006,Porter2009,Fortunato2010}
\begin{equation}
	P^{\mathrm{NG}}_{ij} = \frac{k_{i} k_{j}}{2m}\,,
\end{equation}	
where $k_{i}=\sum_j W_{ij}$ is the strength (i.e., weighted degree) of node
$i$ and $m=\frac{1}{2}\sum_{ij} W_{ij}$. The NG null model is most appropriate for networks in which a
connection between any pair of nodes is possible. Importantly, many networks include (explicit or implicit) spatial constraints that exert a strong influence on which edges are present \cite{barth2011}.  For particulate systems, numerous edges are simply physically impossible, so it is important to improve upon the NG null model for such applications.  We \map{use} the term \emph{geographical constraints} to describe the explicit spatial constraints in such systems.  These constraints exert a significant effect on \map{network} structure, so it is important to take them into account when choosing a null model. For granular materials (and other particulate systems), each particle can only be in contact (i.e., $W_{ij} \neq 0$) with its immediate neighbors. \map{We therefore use a null model,} which we call the \emph{geographical null model}, to account for this constraint. The geographical null model is
\begin{equation}\label{geog}
	P_{ij} = \rho B_{ij} \, ,
\end{equation}
where $\rho$ is the mean edge weight in a network and ${\bf B}$ is the binary adjacency matrix of the network. (Such a null model was used previously for applications in neuroscience \cite{Bassett2013Robust}.) Recall that the adjacency matrix encapsulates the presence or absence of contact between each pair of particles. For a granular material, $\rho = \overline{f} := \langle f_{ij} \rangle$ is the mean inter-particle force. \DB{Because we have normalized the edge weights in the force network ($W_{ij} = f_{ij}/\mathrm{mean}(f)$), we note that in our case $\rho=1$.}

Maximizing $Q$ yields a \map{so-called ``hard partition''} of a network into communities \map{in which} the total edge weight inside of modules is as large as possible relative to the chosen null model. A hard partition assigns each node to exactly one community. (An alternative is a ``soft partition'' \cite{walker-soft}, which allows each node in a network to be associated with multiple communities.) For the geographical null model in (\ref{geog}), maximizing $Q$ \map{assigns the particles into communities} that have \map{inter-module particle forces that are larger than} the mean force. Such \map{communities} represent the force chains in a granular system. 

Because \map{maximizing} $Q$ is NP-hard \cite{Brandes2008}, the success of the maximization is subject to the limitations of the employed computational methods. In the present paper, we use a Louvain-like locally greedy algorithm \cite{Jutla2012}. Additionally, given \map{the numerous near-degeneracies in} the modularity landscape that tends to inflict networks that are constructed from empirical data \cite{Good2010} (i.e., \map{many} different partitions often \map{yield} comparably \map{large values of $Q$}), we report community-detection results that are ensemble averages over 20 optimizations.


\subsection{Properties of Force Chains}\label{props}

We characterize communities using several diagnostics: size, network force, and a \emph{gap factor} (a novel notion that we introduce in the present paper).
The \emph{size} $s_c$ of a community $c$ is simply the number of particles in that community. The systemic size $s$ is given by the mean of $s_c$ over all communities. The modularity $Q$ is composed of sums of magnitudes of bond forces and therefore also has units of force. Therefore, we use the term \emph{network force} to indicate the contribution of a community $c$ to modularity. The network force of a community is given by the formula
\begin{equation}
	\sigma_{c} = \sum_{i,j\in C} [W_{ij} - \gamma \rho B_{ij}]\,,
\end{equation}
where \map{$C$} is the set of nodes in community $c$. The systemic network force $\sigma$ is the mean of $\sigma_c$ over all communities. Communities that correspond to force chains composed of densely packed particles with large forces between them have large values of network force, whereas communities that correspond to force chains composed of sparsely packed particles with small forces between them have small values of network force.

\begin{figure}
\begin{center}
\includegraphics[width=0.8\linewidth]{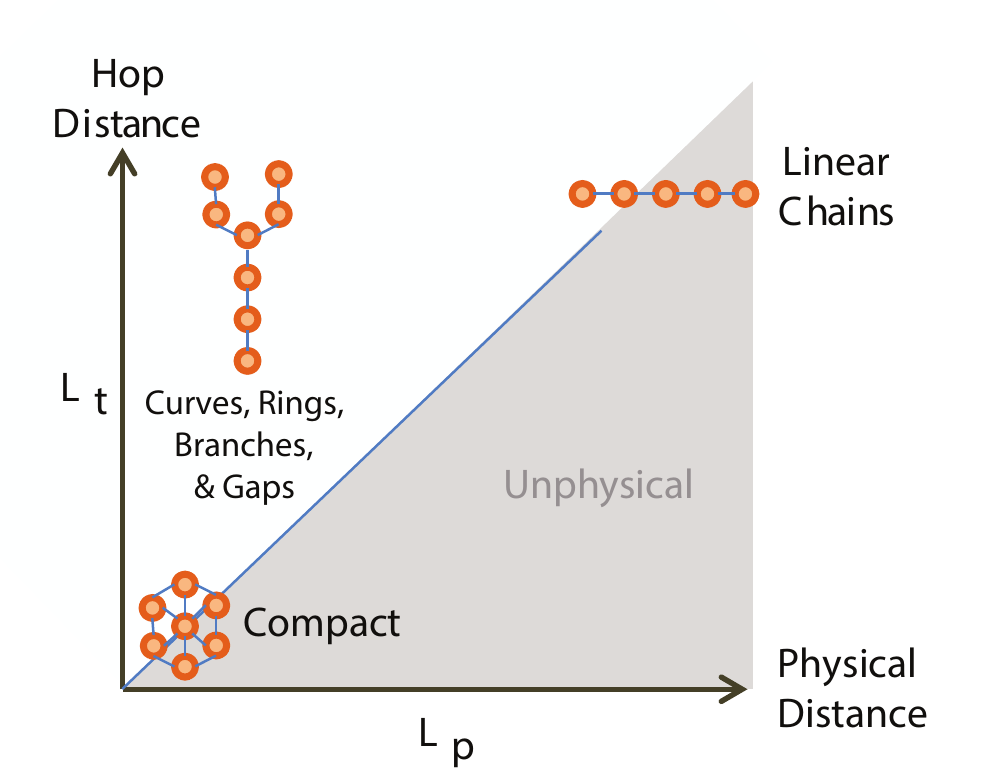}
\end{center}
\caption{(Color Online)
Schematic illustration of the \emph{gap factor}, which we measure via the Pearson correlation between the hop distance $L_t$ and physical distance $L_p$ in granular force networks. Network communities with gaps, branches, and rings can have a larger hop distance than physical distance. They thus reside in the upper triangle.
\label{f:def-gap}}
\end{figure}

To identify the presence of gaps and the \map{extent} of branching in the geometry of a force-chain network, we calculate the Pearson correlation between ``physical distance'' (which we measure using the standard Euclidean metric) and ``hop distance'' (which is often called the ``topological distance'' and counts distance measured only along network edges), and we examine how \map{the correlation} depends on the force-chain topology. \DB{Communities with compact or linear-chain characteristics (see the main diagonal in Fig.~\ref{f:def-gap}) occur when there is a perfect correlation between hop distance and physical distance.} \map{The perfect correlation arises} because particles that are \map{one hop away from} each other (i.e., they are adjacent to each other in the binary contact network) are also 1 particle-distance away. \DB{Small linear chains occur close to the origin, where both hop distance and physical distance are small, whereas long linear chains occur in the upper right quadrant of Fig.~\ref{f:def-gap}. Note that we use the term ``compact'' in the spirit of the mathematical sense of the term, although our exact meaning is somewhat different: ``compact'' communities of particles are at the opposite extreme as linear chains.
Small compact blobs contain particles that are 1 hop-distance away from one another and 1 particle-distance away from one another, and they therefore occur near the origin of Fig.~\ref{f:def-gap}. Larger compact blobs contain particles that are several hop-distances away (and an equal amount of particle-distances away), and they therefore occur in the upper right quadrant of Fig.~\ref{f:def-gap}.}

\DB{In contrast to compact blobs and linear chains,} force chains with gaps, branches, and rings have a larger hop distance than physical distance (see the upper triangle in Fig.~\ref{f:def-gap}). Particles that are close together in space do not necessarily have strong forces \map{to bind} them; the presence of more complicated shapes decreases the correlation between the physical and hop distance. (Note for particulate systems that the lower triangle is unphysical, as it would require forces between particles that are not in contact with one another to achieve a larger physical distance than hop distance.)

\map{To} identify communities that are composed of branched structures, we measure the amount of correlation between the hop distance and the physical distance in the set of all node pairs in a given community. To compute the hop distance of a community, we define the community contact network $\mathbf{B}^{c}$. Its elements $B_{ij}^{c}$ are entries of the matrix $\mathbf{B}$ for which the corresponding nodes have both been assigned to the same \map{community $c$}. We then calculate the path lengths between possible pairs of nodes in a community $c$ using the hop distance on the matrix $\mathbf{B}^{c}$. The resulting matrix of pairwise distances is $\mathbf{L_{t}}$. To compute \map{physical} distance, we calculate the Euclidean distances between \map{all} possible pairs of nodes in a community. The resulting matrix of pairwise distances is $\mathbf{L_{p}}$.

\begin{figure}
\begin{center}
\includegraphics[width=.90\linewidth]{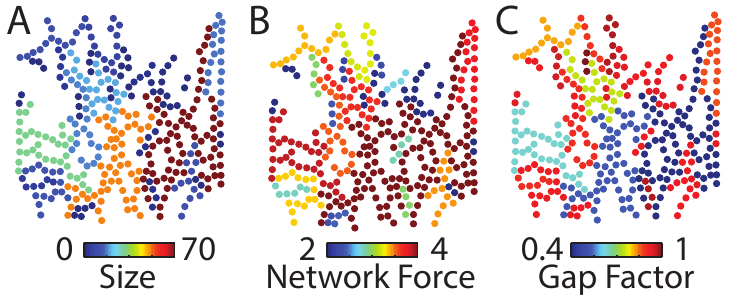}
\end{center}
\caption{(Color Online)
\DB{Sample community-detection results for a single granular experiment at $4.1 \times 10^{-4}E$. We color communities according to \emph{(A)} community size $s_c$, \emph{(B)} network force $\sigma_c$, and \emph{(C)} community gap factor $g_c$. In these calculations, we use $\gamma_{\mathrm{opt}}=0.9$, which we choose to maximize the systemic gap factor $g$ across all pressures (rather than for a single pressure). This example illustrates that the three network diagnostics can reveal very different spatial distributions in the data and thereby supports examining all three diagnostics.}
\label{f:example}}
\end{figure}

In defining the community gap factor, we choose to weight each community by its \map{size. We thus weight large communities more heavily than small communities in linear proportion to the} number of particles \map{that} they contain. In this case, the \emph{gap factor} $g_c$ of a community $c$ \DB{measures the presence of gaps and the extent of branching in a community. We calculate it using with the formula}
\begin{equation}
	g_c = 1-\frac {r_c \, s_c}{s_{\mathrm{max}}}\,,
\label{eq:gapfactor}
\end{equation}
where $r_c$ is the value of the Pearson correlation between the upper triangle of $\mathbf{L_{t}}$ and the upper triangle of $\mathbf{L_{p}}$, and $s_{\mathrm{max}}$ is the size of the largest community. (Note that we exclude the diagonal elements of the matrices $\mathbf{L_{p}}$ and $\mathbf{L_{t}}$.)  \Lisa{To provide further illustration of this quantity, a set of communities colored by their respective gap-factors is shown in Fig.~\ref{f:example}(C).} 

We define the \map{(weighted)} systemic gap factor as
\begin{equation}
	g = 1 - \frac{1}{n} \sum_{c} \frac {r_c \, s_c}{s_{\mathrm{max}}}\,,
\label{eq:gapfactorsys}
\end{equation}
where the quantity $n$ is the total number of communities (including singletons) and we again note \map{that large} communities contribute more than small communities to the averaging in \map{Eq.~}(\ref{eq:gapfactorsys}). An alternative choice for the systemic gap factor is to weight all communities uniformly to calculate an alternative systemic gap factor
\begin{equation}\label{g-uni}
	g_{\mathrm{uniform}} = 1 - \frac{1}{n} \sum_{c} r_c\,.
\end{equation}
As we discuss in more detail in the appendix, branched communities typically include more nodes than linear communities, so the weighted gap factor $g$ tends to be larger for \map{communities with more branching}. By contrast, $g_{\mathrm{uniform}}$ tends to be \map{larger} for more linear communities. For the remainder of the main manuscript, we focus on the weighted gap factor \map{$g$} because it varies far less across packings. Further studies of \map{$g_{\mathrm{uniform}}$} would also be interesting.


\section{Characterizing Force Chains \label{sec:measure}}

We now characterize the force-chain networks of the experiments and simulations that we described in Section \ref{sec:Methods} using the methodology that we described in Section \ref{sec:forcechain}. Our community-detection procedure consists of two stages.  First, we maximize modularity (\ref{eq:one}) with the geographical null model (\ref{geog}) for different values of the resolution parameter $\gamma$: from $\gamma=0.1$ to $\gamma=2.1$ in increments of $\Delta \gamma = 0.2$. We then choose a resolution that approximately maximizes the systemic gap factor $g$ [see Eq.~(\ref{eq:gapfactorsys})]. This ensures that we \map{extract} communities of particles that have strong and dense force-weighted contacts with one another (i.e., \map{our} network partition \map{has a large value of the} modularity $Q$), and that they are spatially sparse (i.e., they tend to have \map{a} \map{small} topological-physical distance correlation $r_c$). The communities that we obtain tend to take the form of chain-like structures 
that are reminiscent of force \map{chains; this provides} visual support that our technique is successful. \map{Our communities are much better than what one obtains} using the NG null model. In previous work using the NG null model\cite{Bassett2012}, we observed that these latter communities always tend to be compact in form.  We know, however, that communities with other qualitative features are very common. (See the schematic in Fig.~\ref{f:def-gap}.)

\begin{figure}
\begin{center}
\includegraphics[width=.90\linewidth]{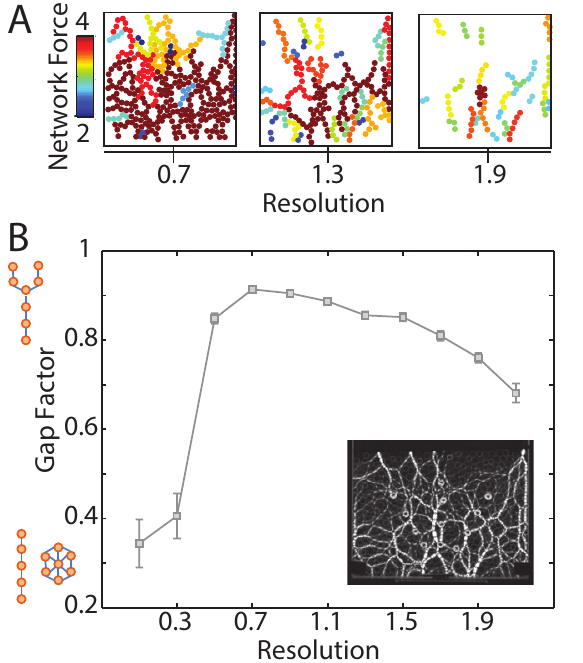}
\end{center}
\caption[]{(Color Online)
Sample community-detection results for a single granular experiment at $4.1 \times 10^{-4}E$ illustrate multiresolution structure in chain-like communities. \emph{(A)} Community structure as a function of the resolution parameter $\gamma$. Color indicates the logarithm (base 10) of the network force $\sigma_{c}$ of community $c$. (We don't show single-particle communities.) \emph{(B)} The systemic gap factor $g$ [see Eq.~(\ref{eq:gapfactorsys})] as a function of the resolution parameter $\gamma$ exhibits a maximum at $\gamma=0.7$ (for $\gamma \in \{0.1,0.3,\dots,2.1\}$). \map{The error} bars indicate the standard deviation of the mean over the laboratory packings.  In the inset, we show an image of the 2D vertical packing of photoelastic disks that we use for panels \emph{(A)} and \emph{(B)}.
\label{f:gamma}}
\end{figure}

The resolution parameter $\gamma$ sets the spatial resolution of the communities \cite{rb2006,Porter2009,Onnela2011} [see Eq.~(\ref{eq:one})]. By tuning $\gamma$, one can either examine large communities (using small values of $\gamma$) or small communities (using large values of $\gamma$). In Fig.~\ref{f:gamma}A, we show an example computation using data from experiments. (Note that we do not show single-particle communities.) Observe that small values of $\gamma$ produce communities that are dominated by compact structures (\map{small} $g$), whereas large values of $\gamma$ produce communities that are dominated by linear structures (\map{small $g$ as well}). This suggests that we can identify an optimal value $\gamma_{\mathrm{opt}}$ \map{that maximizes the systemic gap factor} $g$. This also corresponds to the choice for which the detected communities are most similar to the force-chain structure that we observe visually (see Fig.~\ref{f:gamma}B). For this particular packing, we identify $\gamma_{\mathrm{opt}} = 0.7$ as the \map{best} choice in our 
examination of $\gamma \in \{0.1,0.3,\dots,2.1\}$. As one can see in Fig.~\ref{f:gamma}A, we observe at all values of $\gamma$ that the detected communities vary in their size, network force, and gap factor. We also note that single-particle communities are common for all values of $\gamma$ near 1 (and become increasingly common as $\gamma$ increases after it exceeds 1).

Community detection via modularity maximization with a resolution parameter chosen to yield a maximal systemic gap factor provides an automated approach for detecting force chains in granular media. Once identified, we then calculate the size $s$, network force $\sigma$, and gap factor $g$ to describe key features of a force-chain network. In the \map{following subsections}, we quantitatively examine how the distributions of $s$, $\sigma$, and $g$ vary as a function of pressure for both granular experiments and frictionless simulations.


\subsection{Granular Experiments \label{sec:labpack}}

\begin{figure}
\begin{center}
\includegraphics[width=.90\linewidth]{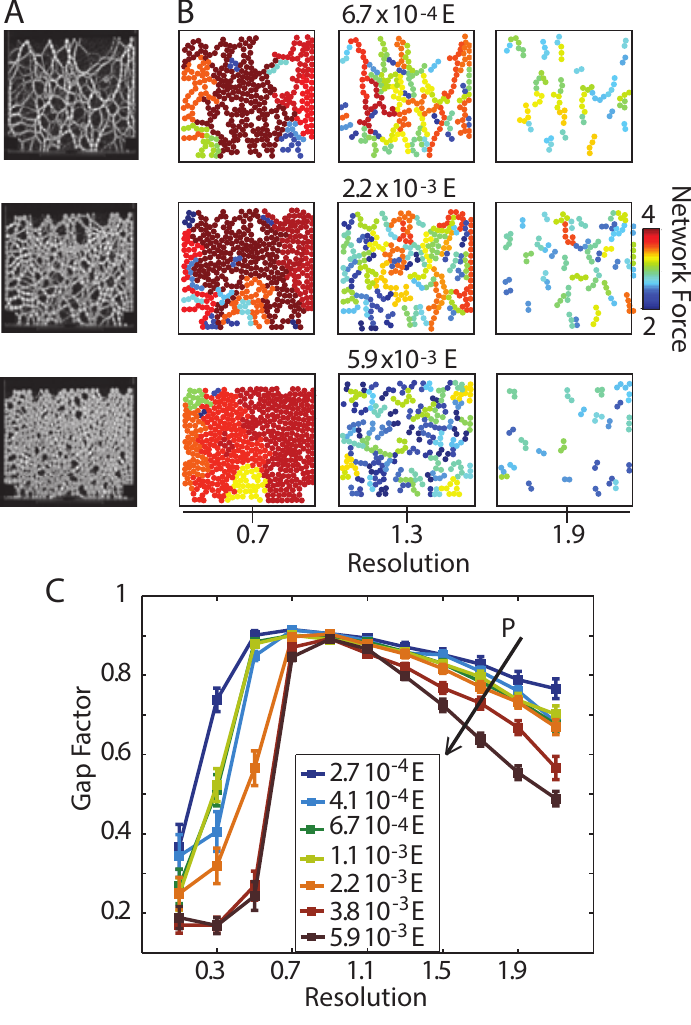}
    \end{center}
\caption{ (Color Online)
\emph{(A)} Images of experimental 2D vertical packings of photoelastic disks. These images reveal the manifestation of the internal stress pattern in a set of photoelastic particles as a network of force chains. \emph{(B)} Community structure as a function of the resolution parameter $\gamma$ for the following pressures: \emph{(top)} $6.7 \times 10^{-4}$E, \emph{(middle)} $2.2 \times 10^{-3}$E,
and \emph{(bottom)} $5.9 \times 10^{-3}$E. Color indicates the logarithm (base 10) of the network force $\sigma_{c}$ of community $c$. \emph{(C)} Gap factor as a function of both the resolution parameter $\gamma$ and the pressure. \map{The error} bars indicate a standard deviation of the mean over
the laboratory packings. The arrow emphasizes increasing pressure.
\label{f:pressureExp}}
\end{figure}

Our experiments are at \map{seven} different pressures, which range from $2.7 \times 10^{-4}E$ to $5.9 \times 10^{-3}E$. As \map{we illustrate} in Fig.~\ref{f:pressureExp}A, we observe that the communities become more compact (so they are closer to the diagonal line in Fig.~\ref{f:def-gap}) as \map{pressure} increases. At a \map{small} value of the resolution parameter ($\gamma = 0.7$), the detected communities metamorphose from branched structures to compact domains as pressure increases. At a \map{large} value of the resolution parameter ($\gamma=1.9$), the detected communities tend to shrink from many-particle chains to $2$-particle chains as pressure increases. This provides a way to quantify our earlier observations that force-chain networks at higher pressures are more homogeneous (observations at small $\gamma$) and less chain-like (observations at large $\gamma$) than they are at lower pressures.

At all pressures (and \map{despite} the aforementioned differences), we find a maximum in the systemic gap factor $g$ as a function of the resolution parameter $\gamma$ (see Fig.~\ref{f:pressureExp}B). The shape of the curve of the gap factor versus the resolution-parameter has a more pronounced peak at higher pressures than at lower pressures: larger slopes descend from and (especially) lead up to the values of $\gamma$ near the maximum. At all pressures, small values of $\gamma$ select compact communities and large values of $\gamma$ select two-particle communities. In between, we observe a value of $\gamma$ at which the most branch-like structures appear; we refer to this value that approximately maximizes the gap factor as the ``optimal'' value. This optimal value changes slightly as a function of pressure; for example, $\gamma_{\mathrm{opt}} = 0.9$ at $5.9 \times 10^{-3}E$ and $\gamma_{\mathrm{opt}} = 0.7$ for $2.7 \times 10^{-4}E$. High-pressure packings also exhibit a much smaller systemic gap factor 
than low-pressure packings at both \map{small} and \map{large} values of the resolution parameter. This observation is consistent with both the increased compactness of the communities and the increasingly homogeneous nature of the force-chain structure as one increases the pressure.

The resolution that maximizes the gap factor identifies structures in a force network that are most reminiscent of the force chains that are apparent by eye; in other words, it identifies branching communities. Therefore, to extract force chains from force networks across different packings and pressures, we \map{examine} community structure \map{for} a range of resolution parameters and identify the resolution-parameter value that approximately maximizes the gap factor. We refer to the communities detected at $\gamma_{\mathrm{opt}}$ as the ``force chains'' in our calculations, and we characterize their properties in terms of their size, network force, and gap factor. \DB{For an illustrative example of a single packing whose communities are color-coded by either size, network force, or gap factor (see Fig.~\ref{f:example}).}

\begin{figure}
\begin{center}
\includegraphics[width=.90\linewidth]{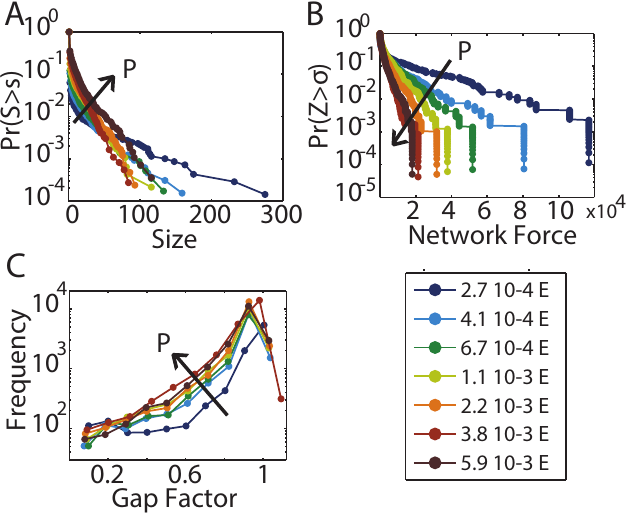}
\end{center}
\caption{(Color Online) Cumulative probability distributions of \emph{(A)} community size $s_c$ and \emph{(B)} network force $\sigma_c$ for all communities. \emph{(C)} Histogram of the gap factor $g_c$ for all communities. In these calculations, we use $\gamma_{\mathrm{opt}}=0.9$, which we choose to maximize \map{the systemic gap factor} $g$ across all pressures (rather than for a single pressure). The arrows emphasize increasing pressure.
\label{f:pressureExpDist}}
\end{figure}

As \map{we illustrate} in Fig.~\ref{f:pressureExpDist}, we observe that the size and network force of the force chains \map{have approximately exponential distributions} for all \map{seven} pressures. This identifies that the majority of communities are relatively small and weak, but a few communities are relatively large and strong. \map{By contrast, the gap-factor} distribution is \map{skewed: most communities have a large gap factor, and only a few communities have a small gap factor}. In high-pressure packings, communities tend to be less broadly distributed with respect to both size and network force; this is consistent with the homogenization of force chains as pressure increases.

\subsection{Frictionless Simulations \label{sec:numpack}}

\begin{figure}
\centerline{
\includegraphics[width=.90\linewidth]{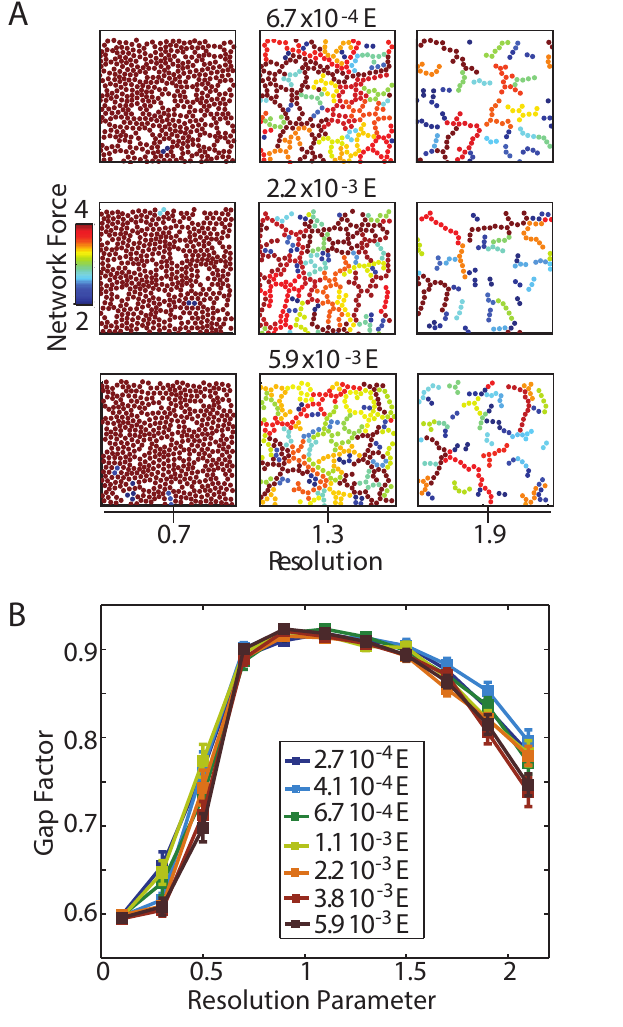}}
\caption{Multiresolution structure of chain-like communities in
numerical packings \map{in frictionless simulations} as a function of pressure. \emph{(A)} Community structure
as a function of the resolution parameter $\gamma$ for the
following pressures: \emph{(top)} $6.7 \times 10^{-4}$E, \emph{(middle)} $2.2
\times 10^{-3}$E, and \emph{(bottom)} $5.9 \times 10^{-3}$E. Color indicates
the logarithm (base 10) of the network force $\sigma_{c}$ of community $c$.
\emph{(B)} Gap factor as a function of both the resolution
parameter $\gamma$ and the pressure. \map{The error} bars indicate the standard deviation
of the mean over the numerical packings.
\label{f:pressureSim}}
\end{figure}

To illustrate quantitative differences in force-chain structure between laboratory and numerical packings, we use our methodology to identify force chains in frictionless simulations. \DB{In addition to their lack of friction, the numerical packings also differ from our experiments in that they use periodic boundary conditions, have zero gravity, have zero fine-scale polydispersity,
 and have a different fabric tensor due to their different initial conditions.} 
 \DB{(A commonality between the two types of packings is that they are both bidisperse.)}
 As in Section \ref{sec:labpack}, we find communities via modularity maximization at different values of the resolution parameter for different packing fractions, which we \map{select} to match the \map{seven} pressures from the experiments.

As \map{we show} in Fig.~\ref{f:pressureSim}A, the simulated packings change qualitatively as a function of the resolution parameter. For relatively small values of the resolution parameter ($\gamma=0.7$), the granular force network has collapsed into a single compact community. In contrast, for relatively large values of the resolution parameter ($\gamma=1.3$), we identify many small branch-like communities that are reminiscent of force chains. Unlike in the granular experiments, we do not observe a strong qualitative change in community structure as a function of pressure (regardless of the value of $\gamma$). Compare Figs.~\ref{f:pressureExp}B and \ref{f:pressureSim}B.

\map{In order to select the most branched community structure, we again identify a value $\gamma_{\mathrm{opt}}$ that is associated with the maximum value of the systemic gap factor $g$.} As with the experimental \map{packing}, there is a clearly identifiable maximum (see Fig.~\ref{f:pressureSim}B), which occurs at $\gamma = 0.9$ for $\gamma \in \{0.1,0.3,\dots,2.1\}$. We also observe that the shape of $g(\gamma)$ is more consistent across pressures for the simulations than it \map{is} for the experiments.

\begin{figure}
\begin{center}
\includegraphics[width=.90\linewidth]{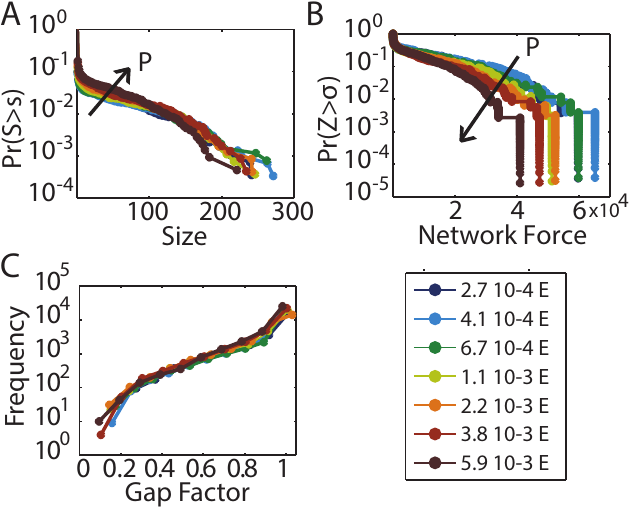}
\caption[] {Size and network force of chain-like communities in numerical packings as a function of pressure. Cumulative distributions of \emph{(A)} community size $s_c$ and \emph{(B)} network force $\sigma_c$ for the resolution-parameter value ($\gamma=0.9$) that maximizes the gap factor in the \map{seven} different pressures separately. \emph{(C)} Histogram of the gap factor $g_c$ 
\map{for} the communities at $\gamma=0.9$. The arrows emphasize increasing pressure.
\label{f:pressureSimDist}}
    \end{center}
\end{figure}

As in the laboratory \map{packings}, we refer to the communities at $\gamma_{\mathrm{opt}}$ as our ``force chains,'' and we characterize their properties in terms of size, network force, and gap factor (see Fig.~\ref{f:pressureSimDist}). Across the \map{set} of pressures from $2.7 \times 10^{-4}E$ to $5.9 \times 10^{-3}E$, we observe that chain size and network force \map{have approximately exponential distributions.} This indicates that the majority of communities are relatively small and weak, but a few communities are relatively large and strong. The \map{gap-factor distribution} has a left skew, which indicates that most communities have a large gap factor and only a few communities have a small gap factor.

In contrast to the laboratory experiments, the shape of the distributions for size, network force, and gap factor in frictionless simulations do not change dramatically as a function of pressure.  Nevertheless, \map{the frictionless simulations to exhibit a systematic difference in the mean size and mean network force for communities as a function of pressure [see Fig.~\ref{f:pressureSimDist}B].}


\subsection{Comparison Between Laboratory and Numerical Force Chains \label{sec:compare}}

\begin{figure}
\begin{center}
\includegraphics[width=.90\linewidth]{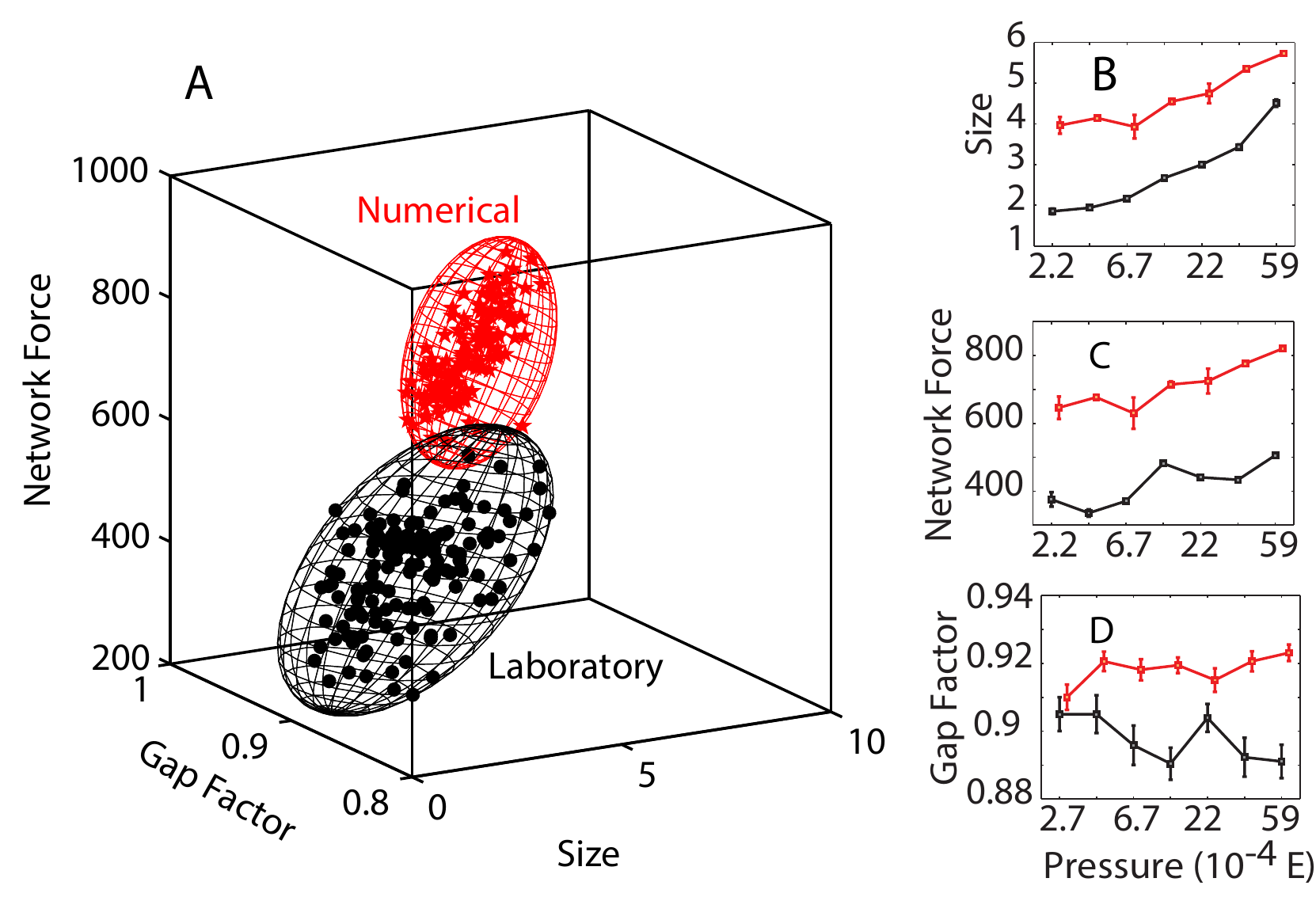}
    \end{center}
\caption{Comparison between the laboratory and numerical packings. \emph{(A)} Scatter plot \map{of} $[s, \sigma, g]$ for each of the 21 runs and \map{seven} different pressures. \emph{(B,C,D)} We average the same diagnostics over all equal-pressure runs for \emph{(B)} size, \emph{(C)} network force, and \emph{(D)} gap factor. All \map{of these calculations use} a resolution-parameter value of $\gamma_{\mathrm{opt}}=0.9$. \map{The error} bars indicate the standard deviation of the mean over the 21 runs.}
\label{f:compare}
\end{figure}

Using our methodology of extracting force chains, it is possible to quantitatively compare and contrast two particulate systems (such the aforementioned granular experiments and frictionless simulations).  In Fig.~\ref{f:compare}, we summarize the three main diagnostics (size, network force, and gap \map{factor)} that we calculate for each of our two case studies.  We anticipate that these diagnostics might provide a helpful means of characterizing how well a given set of simulations---e.g., with anisotropic cell shapes~\cite{Mailman2009,Zeravcic2009} or different models of friction~\cite{Cundall1979, Papanikolaou2013}---captures experimental features.  It also provides a systematic way to compare the properties of different particulate systems or to monitor \map{the temporal evolution of a particulate system.}

The first key difference between \map{our results for the} experimental and numerical packings is that the granular experiments have force chains with \map{smaller} mean size and network force (compare Figs.~\ref{f:compare}A and B). This result is consistent with the more homogeneous nature of force chains in the frictionless packings than in the laboratory packings. Laboratory packings exhibit a few large and strong force chains (see Figs.~\ref{f:pressureExpDist}A \map{and B}), but the majority of a packing tends to be dominated by singleton communities (see Fig.~\ref{f:pressureExpDist}A). By contrast, the frictionless packings have \map{community-size and network-force} distributions that are far less skewed (see Fig.~\ref{f:pressureSimDist}). Additionally, the majority of such a packing is dominated by the force chains (see Fig.~\ref{f:pressureSimDist}A); there are few singletons. These differences in the homogeneity of the packings and in the distributions of size and network force leads on average to 
larger and stronger force chains in the frictionless \map{numerical} packings than in the \map{laboratory} ones.

We also observe that the gap factor is \map{smaller} on average in the laboratory packings than in the frictionless packings. This result is also consistent with the homogenization of structure in frictionless packings, but it might also be influenced by differences between the two types of packings from the presence versus absence of gravity. The frictionless packings are gravity-free, whereas the laboratory packings are influenced by gravity. In contrast to the frictionless packings, the laboratory packings therefore often exhibit linear vertical chains, which decrease the systemic gap factor.  \map{(See our discussion in Appendix 2.)}

Finally, we observe that the size and network force of chains increases with pressure in laboratory packings, whereas the systemic gap factor decreases with pressure for such packings. These observations indicate the presence of a changing length scale, as is to be expected \cite{Wyart2005, Goodrich2013}.
As pressure increases, the force structure becomes more homogeneous, leading to extended sections of material with densely packed particles that exert strong forces on one another. However, the force structure becomes less branch-like, leading to a decrease in the systemic gap factor. The frictionless \map{numerical} packings also exhibit increases in size and network force of chains with pressure; however, unlike in the laboratory packings, the frictionless packings do not exhibit a decrease in systemic gap factor with pressure. This quantitative difference between the two types of packings \map{yields} a technique for measuring the visual differences between Figs.~\ref{f:pressureExp}A and ~\ref{f:pressureSim}A, \map{in which} it is clear that the geometry of the force chains is affected less by pressure in the simulated setting than in the laboratory setting.

Figure~\ref{f:compare}B also demonstrates that the size of \map{communities that resemble force chains} decreases as pressure approaches $0$, which provides some elucidation for a longstanding question in the field of disordered matter. In particulate systems at $0$ temperature, the point at which pressure $P=0$ is the jamming transition, and \map{a lot of} recent work has been devoted to attempting to understand the nature of this transition~\cite{LiuNagelAnnualReview}. There is now strong evidence that there is a {\em growing} length scale~\cite{Wyart2005, Goodrich2013}, which corresponds to the size of regions that are mechanically unstable, near the transition.  As we discuss in Section \ref{sec:Discussion}, it is natural to postulate that our force-chain communities are negatively correlated with weak regions in a particulate packing. Therefore, we expect force chains to shrink as mechanically unstable regions grow.

\section{Conclusions and Discussion \label{sec:Discussion}}

In this paper, we treated granular materials as spatially-embedded networks in which nodes (particles) are connected by edges whose weights we determine from contact forces.  We developed and applied a network-based clustering method, in which we detect tightly-connected ``communities'' of particles via modularity maximization with a geographical null model, to extract chain-like structures that are reminiscent of force chains in both numerical and laboratory packings. From these chain-like structures, we calculated three key diagnostics (size, network force, and gap factor), and we illustrated their utility for identifying and characterizing force-chain network architectures across two types of packings (laboratory and numerical) and across \map{seven} different pressures. To characterize force chains, we identified an ``optimal'' resolution-parameter value that approximately \map{maximizes} the gap factor in \map{each} of these scenarios. An alternative approach would be to examine the structure of force chains that are identified at a set of resolution parameters that one optimizes \map{separately by considering other parameters (e.g., size, network force, etc.).} This is an interesting direction for future study.

After we identified an optimal value for the resolution parameter, we systematically evaluated and compared properties of force-chain communities. A common feature \map{that we observed} in both the \map{(frictional)} laboratory and \map{(frictionless)} numerical packings is that the distribution of force-chain sizes is consistent with an exponential distribution. In the laboratory packings of 2D granular materials, we found that high-pressure force networks exhibit compact rather than branching communities, which is consistent with the notion that increasing pressure causes a breakdown in the long-range heterogeneous structure in a material. In contrast, the geometry of the force chains in the \map{frictionless} numerical packings appear to be affected less by pressure. The force chains in the numerical packings also tend to be both larger and stronger than their counterparts in laboratory packings. Together, these results support the conclusion that the force-chain structure, topology, and geometry is 
different in the two systems.


\subsection*{Methodological Insights}

An important conclusion of our paper is that choosing an appropriate null model is critical for extracting information from community detection in networks. We have developed and applied a \map{geographical null model} for the detection of network communities, which are strongly reminiscent of force chains, in particulate materials. The power of this null model lies in its ability to fix a network's topology (which is given by a binary contact network) while scrambling its geometry (which is given by a force-weighted contact network).  The geographical null model thereby includes more of the fundamental physics of particulate systems than the Newman-Girvan null model that is \map{commonly} used in modularity maximization. An interesting future direction would be to incorporate different physical constraints and principles (e.g., force balance) and to examine the different results that one obtains by including different combinations of relevant physical ideas.  A key question in community detection is what is 
the minimal set of physical ideas---and, more generally, the minimal amount of problem-specific information---that one can include in a null model to get answers that are more insightful than using a one-size-fits-all null model like the Newman-Girvan model.


\subsection*{Practical Utility}

\map{We} expect that our methodology will provide a framework for understanding which features of force chains are universal versus which are governed by detailed particle-particle or particle-environment interactions.
For example, \map{one can use our methodology} to provide a quantification of the similarity of force chains between a simulation and a given experiment.  It can also provide similar quantifications between two experiments (or two simulations) with different types of particles or boundary conditions.  \map{It is also a viable tool to help} predict differences in macroscropic behavior based on subtle differences in the mean or distributions of force-chain diagnostics.

Our methodology also provides additional information that is not available via traditional methods for identifying force chains.  For example, we know which particles are strongly connected within communities, and we can also observe the spatial arrangement of the various communities that comprise a force network.  This allows one to ask whether the arrangement of communities helps to govern the linear and nonlinear response of a disordered packing.  For example, it is possible that large, strong communities indicate a region of relatively high mechanical stability or that boundaries between these communities indicate areas of weakness. In the future, \map{these types of investigations should be helpful for obtaining a better understanding of} the onset of flow or failure in particulate systems.



\section*{Acknowledgements}

DSB was supported by the Alfred P. Sloan Foundation, the Army Research Laboratory through contract number W911NF-10-2-0022 from the U.S. Army Research Office, and the National Science Foundation award \#BCS-1441502. KED and ETO were supported by NSF-DMR-0644743 and NSF-DMR-1206808, and they thank James Puckett for the development of the photoelastic image-processing code. MLM was supported by the Alfred P. Sloan Foundation, NSF-BMMB-1334611, and NSF-DMR-1352184. MAP was supported by the James S. McDonnell Foundation (research award \#220020177), the EPSRC (EP/J001759/1), and the FET-Proactive project PLEXMATH (FP7-ICT-2011-8; grant \#317614) funded by the European Commission. The content is solely the responsibility of the authors and does not necessarily represent the official views of the funding agencies.


\appendix

\section*{Appendix 1: Effect of Community Weighting on Force-Chain Extraction}

\begin{figure*}
\begin{center}
\includegraphics[width=.75\linewidth]{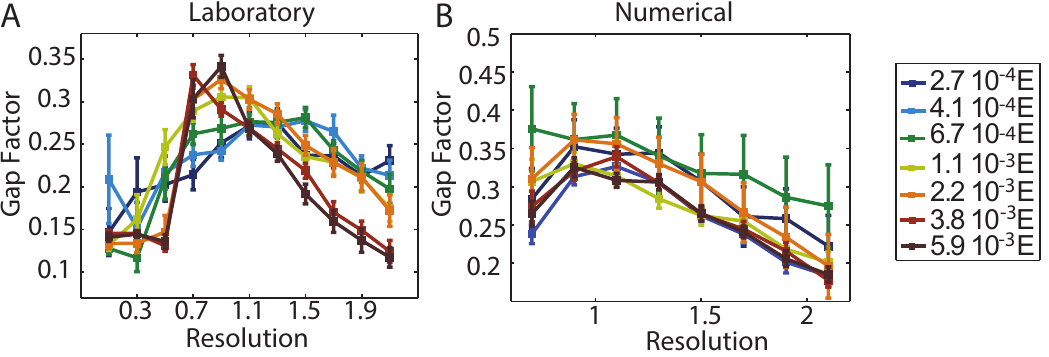}
\end{center}
\caption[]{(Color Online)
\emph{(A)} In the laboratory packings, the systemic uniformly-weighted gap factor $g_{\mathrm{uniform}}$ exhibits a maximum as a function of the resolution parameter $\gamma$ between roughly $\gamma=0.9$ and $\gamma=1.3$ (for $\gamma \in \{0.1,0.3,\dots,2.1\}$ \map{that is different for different pressures.} \map{The error} bars indicate the standard deviation of the mean over the laboratory packings. \emph{(B)} In the numerical simulations, $g_{\mathrm{uniform}}$ exhibits a maximum (which varies little with pressure) as a function of the resolution parameter $\gamma$ at $\gamma=1.1$. \map{The error} bars indicate the standard deviation of the mean over the numerical packings.
\label{f:alt}}
\end{figure*}

In this appendix, we describe an alternative set of measurements that we \map{make} using the \emph{uniformly-weighted systemic gap factor} $g_{\mathrm{uniform}}$, which we defined in Eq.~(\ref{g-uni}).

\map{\subsection*{Gap Factor Versus Resolution Parameter}}

In Fig.~\ref{f:alt}, we observe that the curve of the maximum of $g_{\mathrm{uniform}}$ versus the resolution parameter $\gamma$ varies significantly with pressure in the \map{(frictional)} laboratory packings but not in the \map{(frictionless)} numerical packings. In the laboratory packings, we observe a maximum of $g_{\mathrm{uniform}}$ at $\gamma=0.9$ (for $\gamma \in \{0.1,0.3,\dots,2.1\}$) in high-pressure packings ($5.9 \times 10^{-3}$ E) and at $\gamma=1.5$ for low-pressure packings ($2.7 \times 10^{-4}$ E). In the numerical packings, we observe a maximum of $g_{\mathrm{uniform}}$ at $\gamma=1.1$ for all pressures.
In comparison to our observations in the main text \map{from employing} the size-weighted systemic gap factor $g$, we find that the optimal \map{value of} $\gamma$ is larger when we instead employ $g_{\mathrm{uniform}}$ (compare Fig.~\ref{f:alt} to Fig.~\ref{f:pressureExp} and Fig.~\ref{f:pressureSim}). We also observe that the curves of the systemic gap factor versus resolution parameter exhibit larger variation for the uniformly-weighted gap factor than for the size-weighted gap factor.


\subsection*{Optimal Value of the Resolution Parameter}

The large variation in the maximum of $g_{\mathrm{uniform}}$ over packings and pressures makes it difficult to choose an optimal resolution-parameter value. We choose to take $\gamma_{\mathrm{opt}}=1.1$ because (1) it corresponds to the maximum of $g_{\mathrm{uniform}}$ in the numerical packings and (2) it corresponds to the mean of the maximum of $g_{\mathrm{uniform}}$ in the laboratory packings.
To facilitate the comparison of optimal \map{values of $\gamma$ from} the two weighting schemes, we denote $\gamma_{\mathrm{opt}}$ for $g$ as $\hat{\gamma}$ and we denote $\gamma_{\mathrm{opt}}$ for $g_{\mathrm{uniform}}$ as $\hat{\gamma}_{\mathrm{uniform}}$. Note that $\hat{\gamma}_{\mathrm{uniform}}=1.1$ differs from (and is larger than) $\hat{\gamma}=0.9$.


\subsection*{Force-Chain Structure at the Optimal Value of the Resolution Parameter}

\begin{figure}
\begin{center}
\includegraphics[width=.90\linewidth]{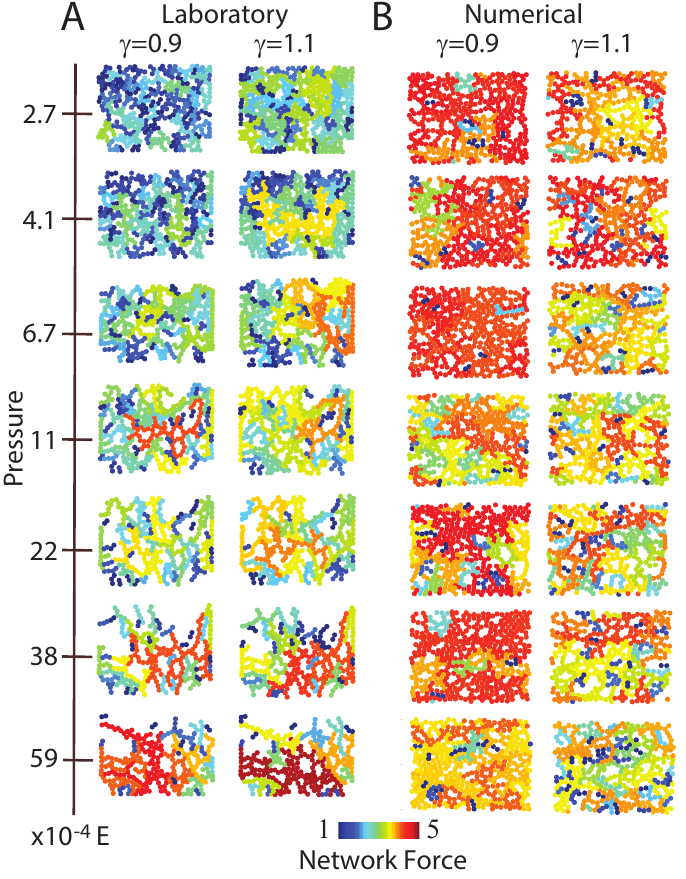}
\end{center}
\caption[]{(Color Online)
In both \emph{(A)} \map{(frictional)} laboratory and \emph{(B)} \map{(frictionless)} numerical packings, we identify larger and more branched force chains at the optimal resolution determined by (\emph{left}; $\gamma=0.9$) the size-weighted gap factor $g$, and we identify smaller and less branched force chains at the optimal resolution determined by (\emph{right}; $\gamma=1.1$) the uniformly-weighted gap factor $g_{\mathrm{uniform}}$. These observations are consistent across all pressure values, but they are especially evident at high pressures in the laboratory packings and are least evident at low pressures in the laboratory packings. \map{In the numerical packings, we observe little variation for different values of pressure.} In both panels, we highlight the network force for the \map{force chains that we identify in} example packings.
\label{f:gammaopt}}
\end{figure}

The force chains that we identify for the optimal value for the uniformly-weighted gap factor (at $\hat{\gamma}_{\mathrm{uniform}}=1.1$) differ from those that we identified in the main text for the optimal value of the size-weighted gap factor (at $\hat{\gamma}=0.9$).  We show our comparison in Fig.~\ref{f:gammaopt}.  For both laboratory and numerical packings, the force chains that we identify at $\gamma=0.9$ are larger and more branched than the ones that we identify at $\gamma=1.1$ (which are smaller and more linear). Indeed, the communities \map{that we identify} at $\gamma=1.1$ have more singletons than the communities \map{that we identify} at $\gamma=0.9$. These results follow from the difference in the two weighting schemes for calculating a systemic gap factor. The size-weighted systemic gap factor $g$ weights larger communities more heavily than smaller ones, and the larger communities tend to be the more branched communities that we identify at smaller values of the resolution parameter (e.g., \map{at} $\gamma=0.9$). In contrast,
 the uniformly-weighted systemic gap factor $g_{\mathrm{uniform}}$ gives equal weight to small and large communities, and it therefore uncovers the linear communities that are evident at larger values of the resolution parameter (e.g., \map{at} $\gamma=1.1$). We can therefore use the size-weighted gap factor to identify larger, more branched force chains and the uniformly-weighted gap factor to identify smaller, more linear force chains.

\section*{Appendix 2: Methodological Considerations}

\subsection*{\DB{Robustness of Community Structure to Errors in the Estimation of Contact Forces}}

\DB{In our frictional laboratory experiments, we estimate that errors in the force measurements could be as large as $\pm 30\%$ of the contact force $f_{ij}$; the high variability arises from the nonlinear fitting process. (Recall that we take particles to be in contact if the force between them is measurable by our photoelastic calculations. We then determine the particle contact forces by solving the inverse photoelastic problem using images taken with polarizers \cite{Puckett2013}.) Somewhat surprisingly, we find that the errors in the force estimates are independent of \KD{both the local force and the} global pressure. To ensure that our results are qualitatively robust to these variations, we construct 20 simulated force networks for each experimental network (21 packings, seven pressures) by adding Gaussian noise with width a $f_{ij}/3$ to each contact. For each of these simulated networks, we reevaluate the estimated community structure (from which we infer the force chains).}

\DB{To determine whether the estimates are robust to this amount of noise, we compare the community structure of the actual force networks with those of the simulated force networks using the z-score of the Rand coefficient \cite{Traud2010}. For comparing two partitions $\alpha$ and $\beta$, we calculate the Rand z-score in terms of the network's total number $M$ of pairs of nodes, the number $M_{\alpha}$ of pairs that are in the same community in partition $\alpha$, the number $M_{\beta}$ of pairs that are in the same community in partition $\beta$, and the number $w_{\alpha \beta}$ of pairs that are assigned to the same community both in partition $\alpha$ and in partition $\beta$. The z-score of the Rand
coefficient comparing these two partitions is
\begin{equation}
	z_{\alpha\beta} = \frac{1}{\sigma_{w_{\alpha \beta}}}
	\left(w_{\alpha \beta}-\frac{M_{\alpha}M_{\beta}}{M}\right)\,,
\end{equation}
where $\sigma_{w_{\alpha \beta}}$ is the standard deviation of $w_{\alpha
\beta}$ (as in \cite{Traud2010}). Let the \emph{mean partition similarity} $z$ denote the mean value
of $z_{\alpha \beta}$ over all possible partition pairs for $\alpha \neq
\beta$.}

\begin{figure}
\begin{center}
\includegraphics[width=.90\linewidth]{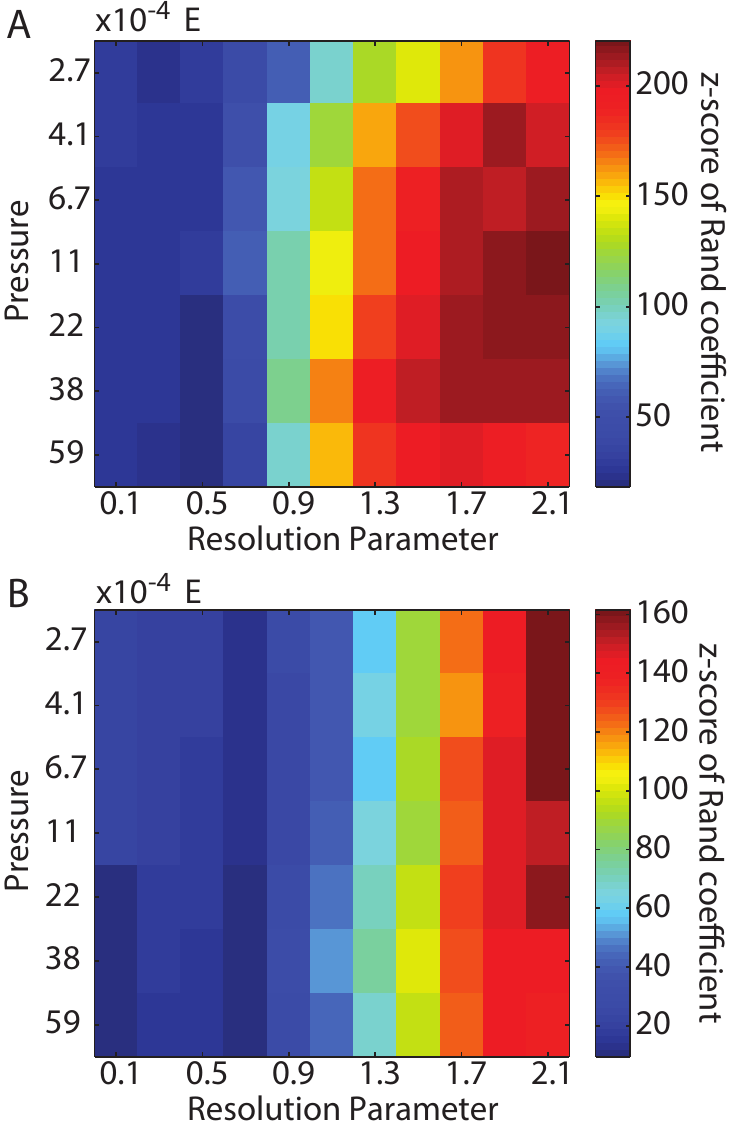}
\end{center}
\caption[]{(Color Online)
\DB{In both \emph{(A)} laboratory and \emph{(B)} numerical packings, we calculate the similarity between the assignment of particles to communities for the actual packings as compared to the packings in which we have added Gaussian noise with width $f_{ij}/3$ to each contact. We estimate partition similarity using the z-score of the Rand coefficient; we compute the z-score for each pair of actual-noisy networks, for each pressure, for each packing, and for each value of the resolution parameter. In this figure, we show the mean z-score of the Rand coefficient over the 21 packings and over the 20 noisy instantiations of the actual force networks.}
\label{f:robust}}
\end{figure}

\DB{We observe that the assignment of particles to communities (and consequently to force chains) in the experimental laboratory force networks is, on average, statistically similar (the z-scores for similarity are larger than 18) to the assignment of particles to communities in the simulated networks constructed by adding Gaussian noise with width $f_{ij}/3$ to each contact (see Fig.~\ref{f:robust}A). These results indicate that our estimates of network communities is relatively robust to the empirical measurement error in the inter-particle contact forces.}

\DB{To further clarify the robustness of our algorithm to variations in inter-particle forces on the order of $\pm 30\%$ of the contact forces, we perform similar calculations for the numerical packings. For each numerical packing, we construct 20 simulated force networks by adding Gaussian noise with width $f_{ij}/3$ to each contact. For each of these simulated networks, we reevaluate the estimated community structure (from which we infer the force chains). We observe that the assignment of particles to communities (and consequently to force chains) in the numerical force networks is, on average,
statistically similar (the z-scores for similarity are larger than 9) to the assignment of particles to communities in the simulated networks constructed by adding Gaussian noise with width $f_{ij}/3$ to each contact (see Fig.~\ref{f:robust}B). These results indicate that our estimates of network communities is relatively robust to measurement error in the inter-particle contact forces in the form of $\pm 30\%$ of the contact force.}

\subsection*{\DB{Effects of Gravity on Community Structure in Laboratory Packings}}

\begin{figure}
\begin{center}
\includegraphics[width=.90\linewidth]{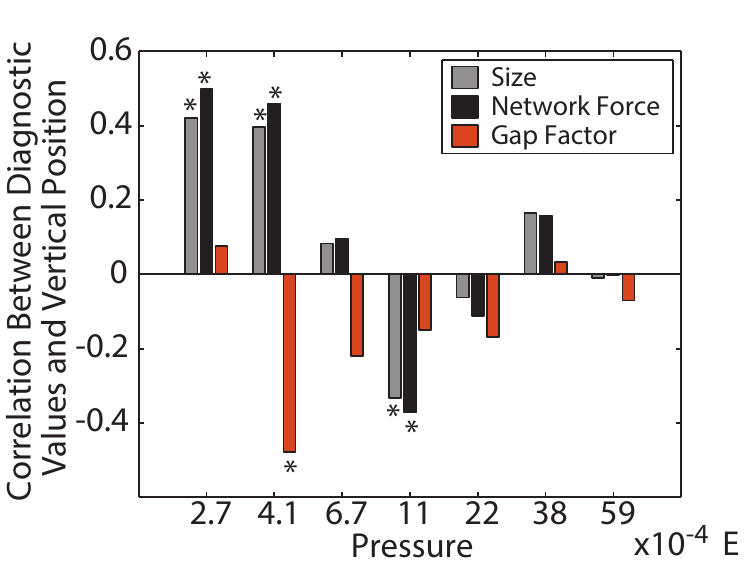}
\end{center}
\caption[]{(Color Online)
\DB{Spearman correlation between network diagnostics (community size, network force, and gap factor) and the mean vertical position of the community as a function of pressure. Asterisks indicate significant correlations at the level of $p<0.05$. (The p-value is uncorrected.)}
\label{f:gravity}}
\end{figure}

\DB{In the laboratory packings, gravity can play an important role in the heterogeneity of observed force chains. To determine the impact of gravity on network diagnostics, we calculate the mean vertical position of particles in each network community for each packing, pressure, and optimization of the modularity quality function for the resolution-parameter value ($\gamma=0.9$) that maximizes the gap factor. For each pressure value separately, we collate these four diagnostics (size, network force, gap factor, and mean vertical position) for every community that we identify over packings and modularity optimizations. To decrease the potential for false positives, we then identify the unique community sizes, and average network force, gap factor, and mean vertical position over all communities of that size. This process serves as a data-reduction procedure. It decreases the degrees of freedom in subsequent statistical testing, and it thereby decreases the potential for false positives. We then ask whether the network diagnostics (size, network force, or gap factor) are significantly correlated with the mean vertical position of the communities. We use a Spearman rank correlation to increase our robustness to outliers in the data. For low values of pressure we observe a significant correlation between each of the three community-level network diagnostics (size, gap factor, and strength) and the mean vertical position of the particles in that community (see Fig.~\ref{f:gravity}).}

\DB{At pressures of $2.7 \times 10^{-4}$E and $4.1 \times 10^{-4}$E, larger communities with larger network force tend to be identified at a lower vertical position of the packing; this is consistent with the compaction effects of gravity. At pressures of $4.1 \times 10^{-4}$E, communities with large gap factors tend to be observed at high vertical positions in the packings (where communities tend to be small and less compact). For values of pressure between $22 \times 10^{-4}$E and $59 \times 10^{-4}$E, we observe no relationship between mean vertical position of particles in a community and community size, network force, or gap factor. These results indicate that gravity can play a role in the shape of force chains, particularly at low pressures.}

\bibliographystyle{apsrev4-1} 
\bibliography{bibfile10,eto4,ked4,lisa}

\end{document}